# Application of remote sensing, GIS and GPS for efficient urban management plan – A case study of part of Hyderabad city

**Akanbi A. K[*], Santosh Kumar[1] and Uwaya Fidelis[2]**

[*,1]Centre for Environment, Institute of Science & Technology, Jawaharlal Nehru Technological University, Hyderabad 50085, A.P, India.

[2]Electronic and Communication Department, Jawaharlal Nehru Technological University, Hyderabad 50085, A.P, India.

## ABSTRACT

*Role of urban planning and management in Hyderabad is becoming more and more crucial due to the dramatic increase in urban population and allied urban problems. Hyderabad is experiencing a rapid urbanization rate. Urbanization contributes many advantages in terms of economics, but if uncontrolled, would produce negative consequences to the physical, social and natural environment. With the advancement of GIS, which considerably influenced the dynamic nature of urban and regional planning, incorporation of GIS becomes imperative for better and improved decision-making in urban planning and management. It offers a solution to the urban problems and decision-making, which is more reliant to the real-time spatial modelling.*

*The integration of Geographical Information System (GIS) and Remote Sensing has provided a tool, which can contribute to much clearer understanding of real planning problems as well as prescriptive planning scenarios to enhance the quality of urban planning and management.*

*Corresponding author: Akanbi A. K.
Plot 28/30, Surulere Estate, Country Home, Ede,
Osun State, Nigeria.
akanbiadeyinka@hotmail.com

## INTRODUCTION

Urbanization is an index of transformation from traditional rural economies to modern industrial one. Kingsley Davis has explained urbanization as process [Davis, 1962] of switch from spread out pattern of human settlements to one of concentration in urban centers. Historical evidence suggests that urbanization process is inevitable and universal.

Urban areas grow in area and population every day, calling for more resources, better living spaces and improved administration. In 1950, only 28 percent of the world population was urban. Today, more than 45 percent of the world stays in urban areas [UNCHS Habitat, 2001]. By 2008, more than half will be living in urban areas, and it is expected by 2030, this figure will cross 60 percent [World Bank, 2005]. Urban areas change in its structure and morphology in varied manner, owing to natural growth as well as the socioeconomic aspirations of the cities.

The urbanization is measured based on the percent urban population and urban-rural ratio. In Table 1 below, it is shown that in the World, about 47 percent population lives in urban areas by 2001. In the More Developed Countries about three quarters of people live in urban areas, in the Least Developed Countries only around a quarter of the population live in urban areas.

**Table: 1 Degree of Urbanization**

| DEGREE OF URBANIZATION IN WORLD, REGIONS, CONTINENTS AND SELECTED COUNTRIES – 2000 | | | |
|---|---|---|---|
| S. No. | Region/Continent /Country | Percent Urban Population | Urban – Rural ratio |
| 1 | World | 47.0 | 89 |
| 2 | More Developed Region | 76.0 | 317 |
| 3 | Less Developed Region | 39.9 | 66 |
| 4 | Least Developed Region | 26.0 | 35 |
| 5 | Africa | 37.9 | 61 |
| 6 | Asia | 36.7 | 58 |
| 7 | Europe | 74.8 | 297 |
| 8 | South America | 79.8 | 395 |
| 9 | North America | 77.2 | 339 |
| 10 | China | 32.1 | 47 |
| 11 | India | 27.8 | 39 |
| 12 | USA | 77.2 | 339 |
| 13 | Indonesia | 40.9 | 69 |
| 14 | Brazil | 81.3 | 435 |
| 15 | Pakistan | 37.0 | 59 |
| 16 | Russian Federation | 77.7 | 348 |
| 17 | Bangladesh | 24.5 | 32 |
| 18 | Japan | 78.6 | 367 |
| 19 | Nigeria | 44.0 | 79 |
| Source: *World Urbanization Prospects – The 1999 Revision – United Nations* Note: For India the data relates to Census 2001 | | | |

### *Trend & degree of urbanization in India*

India shares most characteristic features of urbanization in the developing countries. Number of urban agglomerations / towns has grown from 1827 in 1901 to 5161 in 2001. The trend of urbanization in India is shown in Fig 1.1. India is at acceleration stage of the process of urbanization. The degree or level of urbanization is defined as relative number of people who live in urban areas. Percent urban [(U/P)*100] and percent rural [(R/P)*100 and urban-rural ratio [(U/R)*100] are used to measure degree of urbanization. These are most commonly used for measuring degree of urbanization. The ratio U/P has lower limit 0 and upper limit 1 i.e. $0<U/P<1$.

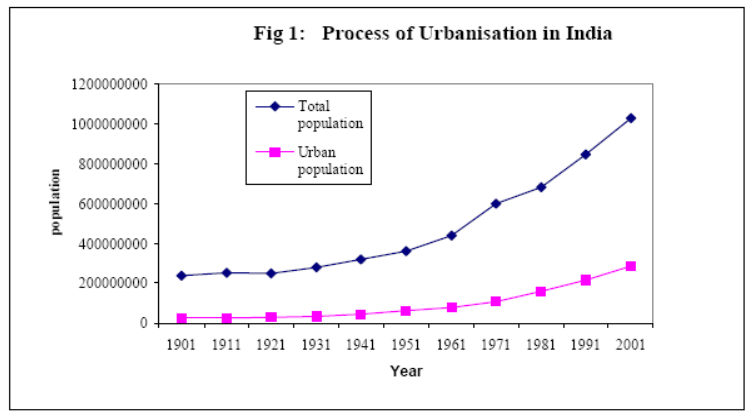

**Figure: 1 Process of Urbanization in India**

The urban-rural ratio in 2001 is about 38, which mean that against every 100 rural areas there are 38 urban areas in India as per Census, 2001. All these indices pinpoint that India is in the process of urbanization and is at the acceleration stage of urbanization as represented in Fig 2.

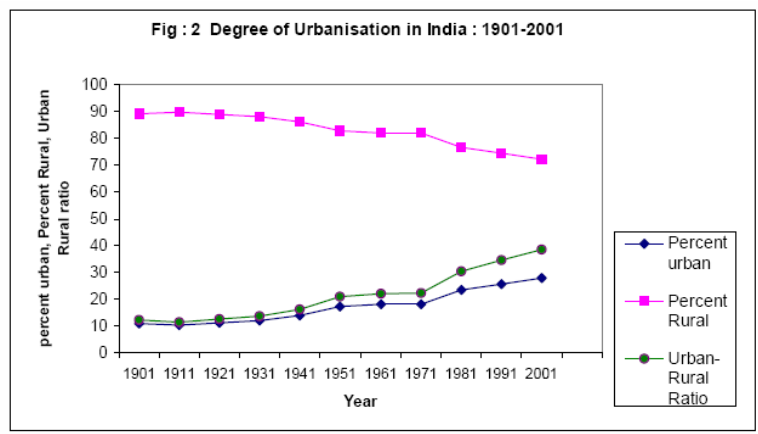

**Figure: 2 Degree of Urbanization in India.**

*Basic feature and pattern of India's urbanization*

Basic feature of Urbanization can be highlighted as:

  i. Lopsided urbanization induces growth of Class I towns;
  ii. Urbanization occurs without industrialization and strong economic base.
  iii. Urbanization is mainly a product of demographic explosion and poverty induced rural - urban migration;
  iv. Urbanization occurs not due to urban pull but due to rural push.
  v. Poor quality of rural-urban migration leads to poor quality of urbanization [Bhagat, 1992].

The big cities attained inordinately large population size leading to virtual collapse in the urban services and quality of life. Large cities are structurally weak and formal instead of being functional entities because of inadequate economic base.

*Problems of urbanization*

Problems of urbanization are manifestation of lopsided urbanization, faulty urban planning and urbanization with poor economic base and without having functional categories:

I. Due to such urbanization, certain basic problems are being witnessed in the fields of:
   a. Housing;
   b. Transport;
   c. Water supply and Sanitation;
   d. Water Pollution and Air Pollution.
II. Megacities grow in urban population [Nayak, 1962] not in urban prosperity, and culture. Hence it is urbanization without urban functional characteristics.
III. Urbanization is degenerating social and economic inequalities [Kundu and Gupta, 1996], which warrants social conflicts, crimes and anti-social activities.
IV. Lopsided and uncontrolled urbanization led to environmental degradation and degradation in the quality of urban life--pollution in sound, air, and water, created by disposal of hazardous waste.

**MATERIALS AND METHODS**

*Application of remote sensing, GIS & GPS*

Our proposed method of solution is an integrated geo-spatial technology i.e. Remote Sensing (RS), Geographic Information System (GIS) and Global Positioning System (GPS) can contribute substantially in a more supplementary fashion to some of the interactive operations that should become an asset for assessing, understanding, mapping utility and service facility using GPS and solving complex urban environmental issues. The objective of this research is to develop a detailed large-scale map and to create computerized user interactive conceptual model for urban activities based thematic maps in an integrated GIS environment using Remote Sensing techniques on GIS platforms

By utilizing remote sensing data and implementing GIS mapping techniques, change detection over a period of time of the urban areas can be monitored and mapped for specific developmental projects. Satellite Remote Sensing, with its repetitive coverage together with multi-spectral (MSS) capabilities is a powerful tool to map and monitor the emerging changes in the urban core as well as in the peripheral areas of any urban areas. The spatial patterns of urban sprawl in all direction over different periods, can be systematically mapped, monitored and accurately assessed from remotely sensed data along with conventional ground data [Lata et-al., 2001]. We therefore describe each component as follows.

**Remote sensing**

Remote sensing has been recognized worldwide as an effective technology for the monitoring and mapping the urban growth and environmental change. The main advantage of satellite remote sensing is its repetitive and synoptic coverage that is very much useful for the study of urban area. It helps to create information base on land use, land cover distribution, urban change detection, monitoring urban growth and urban environmental impact assessment. Satellite images enable us to better understand some of the intrinsic components of urban ecosystems and the interactions within whole urban environment. Remote Sensing technology can be put to best use if it is incorporated with GIS [Longley, 1999].

*Geographical information system (GIS)*

GIS is basically an information system that deals with spatial data. As almost all municipal data has spatial relevance GIS assumes a central role in such a system. Geographic

Information System is a computer-based system to capture store, edit, display, and plot geographically referenced data. Geographic Information System provides for input, co-ordinate registration/transformation, management, query, analysis, modelling, map composition and production of cartographic & maps.

GIS does not hold maps or pictures – it holds a database. The database concept is central to GIS and is the main difference between a GIS and a computer mapping system, which can only produce good graphic output. GIS thus incorporate a data base management system.

The advantages of Geographic Information System in data handling and inferencing. The integrated framework of Remote sensing techniques and GIS framework greatly reduces time, effort and expenses in using geographical data.

*Global positioning system (GPS)*

The explosion in interest in GIS as the management's tool has been accompanied by the development of number of enabling technologies, one of the most important of which is the GPS (Global Positioning System) [Lange, A.F and Gilbert, C, 1999].

GPS advantages are:

i. GPS may be used to identify or define the geographical co-ordinates associated with satellite imagery. GPS is used to reduce distortions and to improve the positional accuracy of these images.
ii. GPS can be used in the ground truthing of satellite images.

*Proposed system architecture*

Our proposed architecture involves the use of remotely sensed satellite images and GPS data in a GIS system. In this architecture, super imposing the satellite images on the respective GIS data, this give perspective urban information where inference can be drawn about future urban trends, easily decimates information.

*Implementation of the proposed architecture: the case study*
*An overview of the study area (Hyderabad)*

Hyderabad, the capital of Andhra Pradesh State is located in the heart of Deccan Plateau and lies approximately at 17°- 21N latitude and 78°- 30E longitude. The city is bounded by Rangareddy and Medak districts. The urban growth of the city has taken place rapidly to the northeast and northwest and has spread beyond the Municipal limits. The river Musi divides old and new Hyderabad while the former is located on southern bank and latter is located on northern bank.

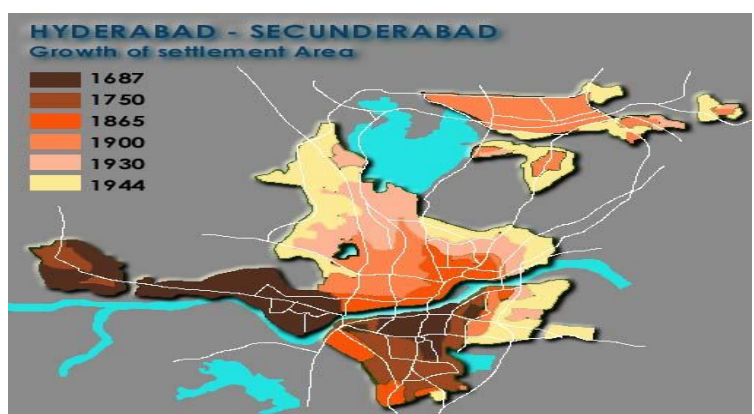

**Figure: 3 Hyderabad growth along the years.**

Table: 2 Population growth in Hyderabad

| Population Growth in Hyderabad 1901- 2001 |||
|---|---|---|
| Year | Total Population | Annual Gr. Rate (%) |
| 1901 | 499,082 | |
| 1911 | 627,720 | 2.32 |
| 1921 | 556,913 | -1.19 |
| 1931 | 588,217 | 0.55 |
| 1941 | 810,790 | 3.26 |
| 1951 | 1,083,634 | 2.94 |
| 1961 | 1,191,668 | 0.95 |
| 1971 | 1,682,284 | 3.51 |
| 1981 | 2,251,009 | 2.96 |
| 1991 | 3,145,939 | 3.4 |
| 2001 | 3,686,460 | 1.6 |

*Density*

Hyderabad District, with a density of 16,988 persons per sq. km. as per the Census of India 2001 is the district with the highest density in the state of Andhra Pradesh, being almost fully urban. The average density for the state (which includes rural areas) is much lower at 275 persons per sq. km. in 2001.

*Methods of data acquisition*

Depending on the availability of maps for the study area, features for the Base Map and availability of other sources of such data products are derived and extracted through various sources, which are given in the following.

Table: 3 Data Acquisition

| Type of Data | Source of Data |
|---|---|
| Toposheets (1:50,000 Scale) | SOI (Survey of India), Government of India. |
| Satellite Data (IKONOS DATA) | NRSC (National Remote Sensing Centre) |
| Meteorological Data | Indian Meteorological Department (IMD) |
| Maps showing<br><br>Existing information of Hyderabad and its boundary of GMCH area | Municipal Corporation of Hyderabad (GHMC)<br><br>Centre for Environment, JNTUH.<br><br>APPCB<br><br>APSRAC |
| Software used<br><br>AutoCAD, ArcInfo, ArcView, ArcGIS 9.1, Erdas 8.6 | Licensed Versions from ESRI – New Delhi |
| Field Data | Through intensive field work<br><br>(Ground truth) |

*GIS data types*

Basically all the GIS data used in this study are classified as

1. Topographical data
2. Thematic data
3. Collateral data

The topographical and thematic data are classified as spatial data and the collateral data as attribute data. The details of these types of data products are discussed below.

*Generation of topographic maps*

Raw geographical data are available in many different analogue and digital forms such as toposheets, aerial photographs, satellite imageries and tables. In this research, the base layers generated from toposheet are:

(i)     Base map
(ii)    Drainage map
(iii)   Transportation Network map
(iv)    Watershed map
(v)     Slope map

These paper-based maps are then converted to digital mode using scanning and automated digitization process. These maps are prepared to a certain scale and show the attributes of entities by different symbols or coloring. This entire process is geo-referenced. The same procedure is also applied on remote sensing data before it is used to prepare thematic maps from satellite data.

*Methodology of the workflow*

Remote sensing techniques and GIS tools have become important in management of urban environment. A number of studies have demonstrated this [Rashid and Sokhi, 1995; Sokhi, 2001; Subudhi, 2001; Roy, 2002]. The main advantage of satellite remote sensing for the monitoring and mapping the urban growth and environmental change, is its repetitive and synoptic coverage that is very much useful for the study of urban area. It helps to create information base on land use, land cover distribution, urban change detection, monitoring urban growth and urban environmental impact assessment.

The important one of the basic elements of GIS is the data. In collecting data there are various methods of them, GPS or classical measurement, Aerial Photographs, Satellite Images and Digitizing of Map and Documentary are the principle methods. In GIS, data have been obtained from different resources by using software, hardware and technology.

One major part of the GIS is the ability to overlay various layers of spatially referenced data, which allow the users to determine graphically and analytically just how structure and objects. e.g. roads, water distribution and community zoning) interact with each other.

*Survey methods*

A comprehensive physical survey of the planning area can be carried out using total station survey equipment. Several details are surveyed ranging from all built features, roads, natural elements, levels etc.

*Satellite images*

Availability of satellite images has made the task of correcting and updating the existing city maps much easier and faster. A reasonably accurate Base Map can be prepared using the satellite images (PAN and IKONOS) as a base and integrating information from various sources such as aerial photographs, revenue maps, SOI Topo-sheets, maps from various departments etc. Appropriate corrections are required to ensure geographical accuracy such as geo-referencing and registration of satellite images with topographical sheets

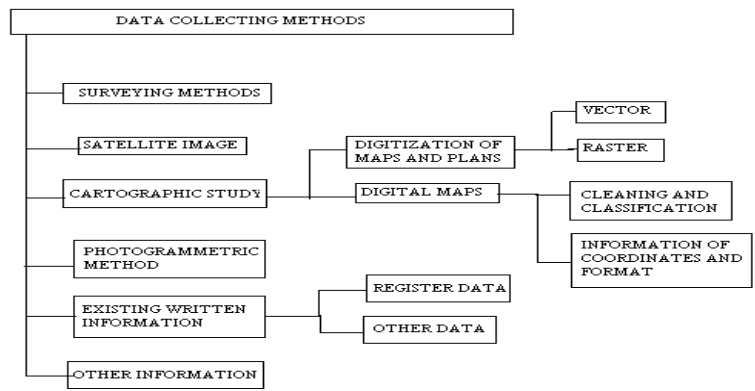

**Figure 4: Workflow Methodology**

*Methodology for data converstion (preparation of base map):*

*Data acquisition*

The different maps (e.g. IKONOS map / aerial photos) of different scale containing the different theme of the study area are collected in digital format and further collated to evolve the strategy for data integration and attribute data capturing into different theme layers.

*Data evaluation*

The data evaluation for each map collected shall be done keeping in view the source and its reliability and also positional accuracy.

*Data processing*

Data processing is done under these stages: -
- *Pre-processing*: This step involves making the data ready and suitable for scanning.
- *Scanning*: The maps in hard copy form shall then be scanned with the most appropriate threshold values.
- *Warping/rubber sheeting*: At this stage the data is transformed to exact dimensions through software, which normally uses polynomial transformation.
- *Raster Editing*: In this operation all the speckles / smudging etc. are deleted and data is cleaned.

*Database structure and design*

At this stage the design of database shall be prepared for the required objectives, which shall include the following:

- Design of layout of sheets for all the area i.e. indexing;
- Structuring of layers of storage of different items of detail;
- Generation of Symbology.

*Digitization / Vectorization*

The digitization is done on the screen using semi-automatic line following techniques under 'zoom in' environments to achieve better accuracy following standard Survey of India norms.

The Digitization process shall involve:

- Vectorization: Conversion of raster data to vector form.
- Symbolization: Generation and attachment of cartographic symbols.
- Layering: Layering of data as per the needs.
- Symbol Coding: Codification of symbols in a uniquely classified manner.
- Edge Matching: Each edge is matched / adjusted individually item by item. This is necessary to have a seamless database.
- Attribute Attachment: Attributes are attached pertaining to each of the relevant category /category.
- Data Base Linking: Linking of the database to each property for whole of the town.

*Ground survey for geo-coding and geo-referencing*

**A.** *Ground Control*

Ground Control Verification are done using Total Stations covering the periphery of the study area and providing control points of whole to the part. The control point is used for Geo-coding and Geo-referencing of the digitized maps.

**B.** *Updating by Satellite Data*

Using satellite Imagery the digitized and updated using NRSA data.

**C.** *Ground Validation*

Once the maps have been digitized and updated, ground validation is carried out.

**RESULTS AND DISCUSSION**

*Methodology integrated information system*

The vector data files are stored in the local hard disk in the ESRI's Shape files and the attribute information in the database. The application software integrates the vector data with the access data for analysis purposes.

*Generation of spatial & Attribute database*

**PROCEDURE FOR PREPARING THEMATIC MAPS**

Procedures for data validation are incorporated into the basic software system with automatic data processing. Satellite imageries are geo-referenced using the Ground Control Points with SOI toposheets (1:25,000) as a reference and further merged using ERDAS (8.6) software to obtain a fused high resolution IKONOS data (1- Meter Resolution). The study area is then delineated and subsetted from the fused data based on the latitude and longitude values and the final hard copy output is prepared for the generation of thematic maps using visual interpretation techniques. Spatial Database like thematic maps like Base map, Transport network and Drainage network maps are prepared from the SOI toposheets on 1:25,000 scale using ARCGIS (9.1) software to obtain a baseline data. Thematic maps of the study area were prepared using visual interpretation technique from the fused satellite imageries and SOI toposheets along with ground truth analysis.

*Base map*

Base map was prepared from SOI toposheet no 56K/7 NE on 1:25000 scale which was overlaid on satellite imagery, IKONOS data. To get an accurate ground control points deletion of certain features like road network, water bodies, canals settlements etc the toposheet are used for exact matching with those on the satellite imagery. This leads to preparation of the base map. The base map showed following features like settlements, which are categorized into dense, medium, sparse etc. The other features like major water bodies, major road network, drainage, pattern etc.

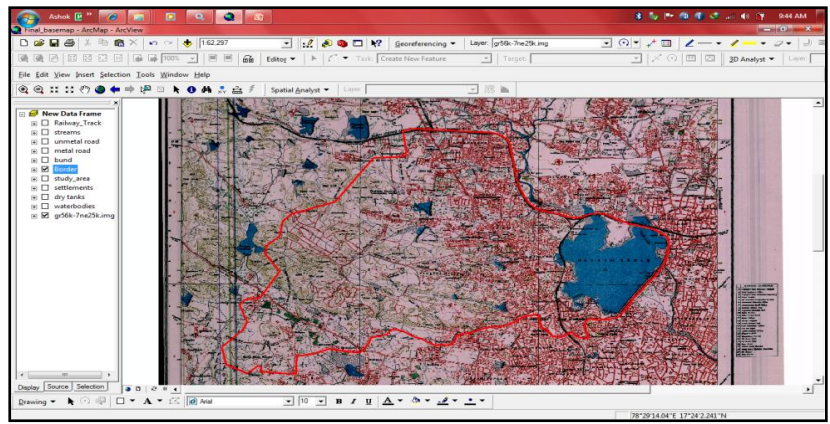

**Figure: 5 Subsetting of the Study Area**

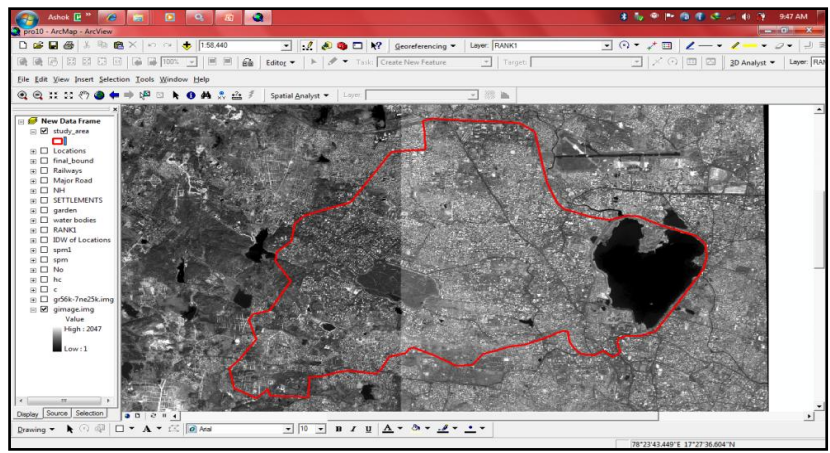

**Figure: 6 Delineating the Study Area**

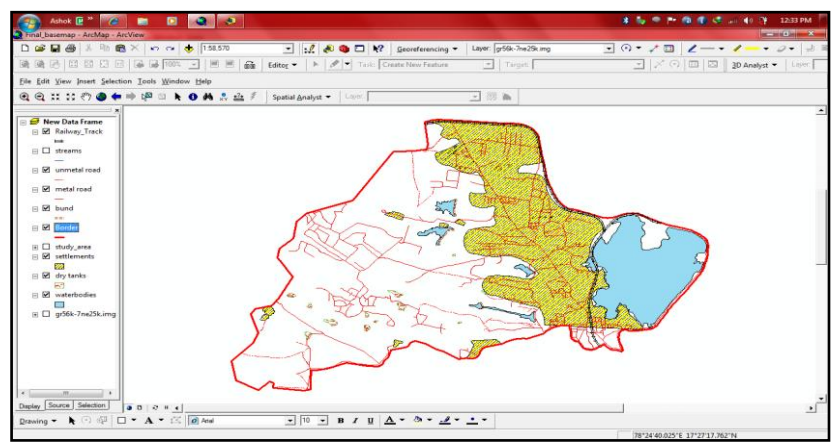

**Figure: 7 Base Map of the Study Area**

*Drainage map*

Drainage network was prepared from SOI toposheet no 56k/ 7 NE of the respective study area with the help of base map prepared. The major water bodies present in this zone are Durgam cheruvu, Hakimpet cheruvu Yousufguda cheruvu, Sheikpet cheruvu etc. The slope direction can also be known through drainage network map, which is useful in understanding topography, geomorphology, soil type and its erodability etc.

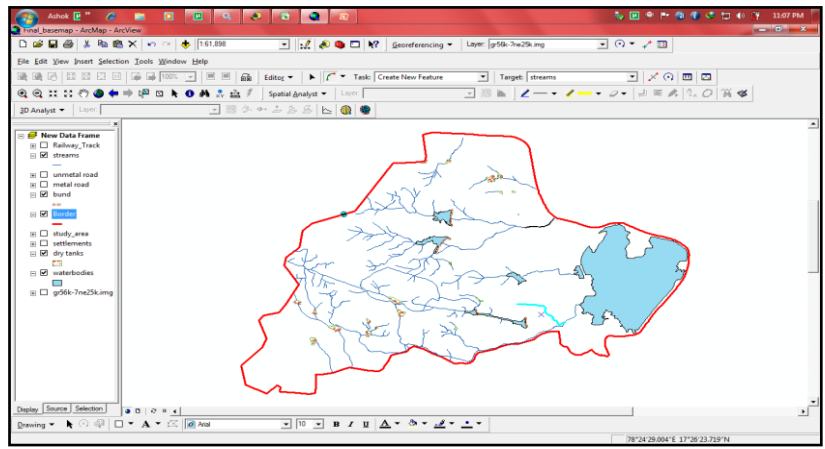

**Figure: 8 Drainage Map of the Study Area**

*Road network map*

Road network was prepared from SOI toposheet no 56k/7 NE of the respective study area with help of base map prepared. This showed the major roads passing through this zone, which includes NH-9 to Mumbai, which is passing from khairatabad to Sanath Nagar in this zone. And various other roads connecting the settlements, landforms etc. which was very help full during sampling.

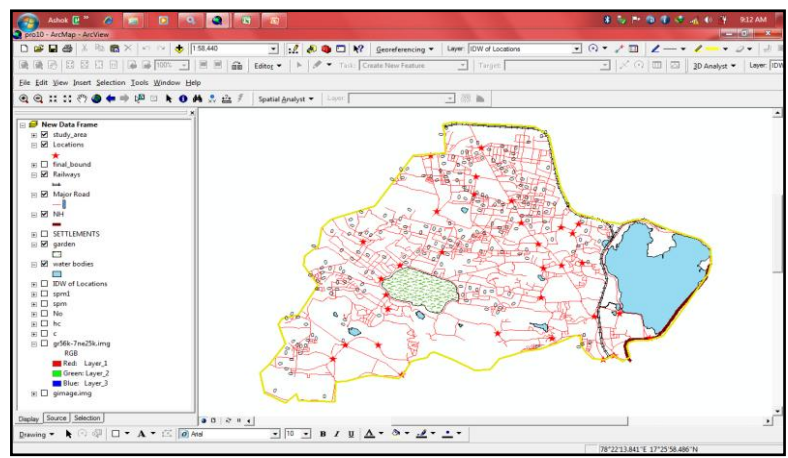

**Figure: 9 Road Network Map of the Study Area**

*Master plan approach and development of urban attribute data*

Urban planning is basically resource generation, resource development and resource management exercise. The efficiency of urban settlements largely depends upon how well they are planned, how economically they are developed. Planning inputs largely govern the efficiency level of human settlements. Urban planning and development refers to a process that harnesses spatio-economic potential of an area for the benefit of the people. Urban planning includes preparation / renderings are:

1. Perspective Plan
2. Development Plan
3. Annual Plan
4. Schemes and projects
5. Participatory approach for supply of land and infrastructure development.

*The master plan approach*

The master plan, which was perceived to be a process rather than a conclusive statement, provides guidelines for the physical development of the city and guides people in locating their investments in the city. In short, Master Plan is a design for the physical, social, economic and political framework for the city, which greatly improves the quality of Urban Governance also.

The functions of the Master Plan / Development plan are as follows:

i. To guide development of a city in an orderly manner so as to improve the quality of life of the people
ii. Chart a course for growth and change, be responsive to change and maintain its validity over time and space, and be subject to continual review
iii. Direct the physical development of the city in relation to its social and economic characteristics based on comprehensive surveys and studies on the present status and the future growth prospects; and

The aim should be to make urban planning system as a continuous process. Each level of plan must include measures for infrastructure development and environmental conservation:

i. *Perspective Structure Plan:* The long-term Perspective Structure Plan could be prepared by the MPCs broadly indicating goals, policies and strategies for spatio-economic development of the urban settlement.
ii. *Infrastructure Development Plan*: Integrated infrastructure Development Plan should be prepared by ULBs in the context of the approved Perspective Plan. The scope of the Plan should cover an assessment of existing situation, prospects and priorities and development including employment generation programs, economic base, transportation and land use, housing and land development, environmental improvement and conservation programs
iii. *Annual Plan*: The Annual action plan should provide and in-built system for implementation of the Development Plan. In this plan various urban development schemes should be integrated spatially and financially.
iv. *Projects and Schemes*: As part of the Development plans and Action plans, projects and schemes within towns / cities could be taken up for any area / activity related to housing, commercial centers, industrial areas, social and cultural infrastructure, transport, environment, urban renewal etc. by governmental bodies / local agencies / private sector and through public private-partnership. Such projects could be both long-term and short-term and in conformity with the development requirements of the respective town / city.

## CONCLUSION

We have shown in this paper that in order to achieve healthy living conditions in our urban areas it is necessary to resort to innovative and efficient Urban Management Plan (UMP) using application of remote sensing, GIS and GPS into an Integrated Information System, which have to play an important role not only in fighting the urban growth, but used to minimize the many negative effects of urban growth, such as traffic problems, slums and environmental degradation and aid in decision-making by providing data which are utilized for accurate and correct assessment.

## ACKNOWLEDGEMENT

The authors wish to acknowledge the support rendered by Survey Of India, National Remote Sensing Centre, Indian Space Research Organization, and Centre for Environment, Institute of Science & Technology, Jawaharlal Nehru Technological University, Hyderabad, A.P, India## REFERENCES

1. Anji Reddy.M, 2001 textbook of RS and GIS, second edition, B.S publications, Hyderabad.
2. Anji Reddy M, 2005, "Environmental Geoinformatics and Modeling"; Proceedings of International Conference on Environmental Management, B.S. Publications.
3. Batty M, 1992, Urban modeling computer graphic and geographic information systems environment, Environment and Planning Board.
4. D.P.Tiwari, Challenges in Urban Planning for local bodies in India, I.A.S., Commissioner & Director, Town and Country planning, Bhopal.
5. Dr.S.P.Sekar 2000, GIS Application for Urban Planning – A Case study of Tinidivanam Town, Tamil Nadu, GIS Development.
6. Encylopaedia of Environmental Sciences 1992, revised volume 15.
7. Govt. of India, 1988, Report of National Commission on Urbanization, Volume – VI, Published by Govt. of India.
8. G.K.Tripathy, Urban Planning and information system for Municipal corporations, Tata Infotech Ltd, Mumbai.
9. Municipal Geographic Information Systems 2001, A note to the Empowered Committee (EC), APUSP/Governance & Reform component/Municipal GIS/Note to EC/10 Nov 21$^{st}$.
10. Manual of Land use / Land cover mapping using Satellite imagery", Part I and II 1989, National Remote Sensing Centre, Department of Space, Govt. of India.
11. National urban Information System 2008 (NUIS), Manual for thematic mapping using high resolution satellite data and geospatial techniques by National Remote sensing Agency, Urban studies & Geoinformatics Group, dept. of space, Govt of india, Hyderabad.
12. NRSA, 1994, Mapping and Monitoring Urban sprawl Hyderabad city. Project report, 1-84.
Novus International Journal of Engineering & Technology 2013, 2(4)      13